# A simple way of simulating transmission lines using free software. Application to a quarter wave transformer.

F. A. Callegari

Centro de Engenharia, Modelagem e Ciências sociais Aplicadas, CECS, Universidade Federal do ABC, Santo André, 09210-170, SP, Brazil. e-mail: fulvio.callegari@ufabc.edu.br

*Abstract*—**A simple way of simulating transmission lines using a free software is presented. In order to validate the results, the effective value of reflection coefficient for voltage wave in a quarter wavelength line was determined trough a transient analysis, for an incident voltage sin-type continuous wave front. The simulations presented reproduced the results previously found analytically.**

## I. INTRODUCTION

In this work, a simple method of simulating transmission lines is presented, in such way some basic concepts may be better apprehended by the student of this subject. To validate the results achieved, theoretical results found in [1] were verified by our simulation. In order to facilitate the lecture of this work, I first summarize the principal results obtained in [1], then, the simulation is presented. The results obtained by the simulation are in agreement with the previously calculate ones.

Transmission line may be defined at a basic level as the device that transmits electromagnetic energy between two points in a controlled way. This transmission is guided trough a physical medium formed by two conductors separated by an insulator material, from a source energy generator to a load. Ideally, all of the energy generated at the source would efficiently reach the load and be consumed by it. Unfortunately, this is hardly the case. The transmission line always suffers from loses due to the attenuation of the conductors, a topic that will not studied here. Also, the energy reaching the load is generally not totally absorbed, producing unwanted reflections. The energy in a line is transmitted through voltage and current waves. Therefore, an energy reflection means that voltage and current waves are returning back to the generator. It is well known that when two counter propagating waves interact, the result is a stationary wave pattern, which has maximum and minimum amplitude values. These extremes values remain at the same point in space. A stationary voltage wave in a transmission line may origin some problems; one of them is that the maximum voltage value of the wave, which is the sum of the incident and the reflected amplitudes waves, may damage the insulator material trough dielectric breakdown, producing a short circuit in the line. Also, the energy returning back could damage the generator. For these reasons, methods to avoid energy reflections in transmission lines are important topics in textbooks on electric engineering. When these reflections are absent, or were avoided using some technique, it is said that the line is matched to the load.

Now, a brief introduction to the basics concepts on transmission lines is presented. In figure 1 a basic schematic is shown.

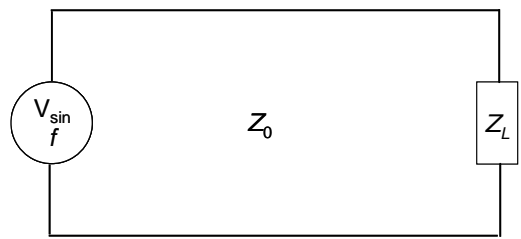

Fig 1. Schematic showing energy generator, transmission line and load

There will be considered sin-type voltage waves, characterized by the amplitude $V_0$, and the frequency $f$. With the knowledge of the electric permittivity $\varepsilon$, and magnetic permeability $\mu$, of the dielectric (insulator) medium of the line, the wavelength $\lambda$, corresponding to the voltage wave can be determined by $\lambda = 1/f(\varepsilon\mu)^{-1/2}$ [2]. The quantity $Z_0$ is the characteristic impedance of the line, which is defined as the impedance calculated as the quotient between the voltage and the current propagating by the line, taking account the waves in just one sense of propagation. The quantity $Z_L$ is the load impedance. In this work, there will be considered lines without looses, i.e., $Z_0$ real.

The reflection of the voltage wave at the load can be quantified by the reflection coefficient of voltage, defined as the quotient between the amplitude of the reflected to the incident voltage wave [3], given by:

$$\rho_V = \frac{V_R}{V_I} = \frac{Z_L - Z_0}{Z_L + Z_0}. \tag{1}$$

Where $V_R$ and $V_I$ stands for the amplitudes of voltage waves reflected and incident, respectively, measured at the load.

It can be seen from Eq. (1), that unwanted energy reflections are due to the dissimilar values of the characteristic impedance of the line and the load impedance.

In most of the textbooks treating on transmission lines, the quarter wave line, or quarter wave transformer, is presented as one of the main devices used to avoid unwanted energy reflection from a load.

The characteristic impedance of such a line is given by [1]:

$$Z_{0\lambda/4} = \sqrt{Z_0 * Z_L}. \tag{2}$$

The physics length of this line is $\lambda/4$.

This line must be connected between the transmission line and the load. Also, this line can be used to match the impedance between two lines of different characteristics impedances.

By the other side, the characteristic impedance of the transmission line, $Z_0$, is different from $Z_{0\lambda/4}$. Applying Eq. (1) to the junction point between the transmission line and the quarter wavelength line, it can easily seen that the value of $\rho_V$ will not be zero and, at a first sight, the original purpose of the quarter wavelength line would not have been fulfilled, i.e., there will be reflected voltage waves returning back to the generator. Textbooks usually used in undergraduate courses [2]-[4] do not comment about this apparent contradiction when the quarter wave line is presented.

One of aims of this work is to clarify this point trough a transient analysis, taking account partial voltage reflections and transmission at the junction point between the transmission line and the quarter wavelength line, and considering also the accumulated phase shift of the voltage wave due to propagation trough the line. This analysis is applied to a particular example of a quarter wave line, given in ref [3]. There, the treatment of such example is fairly intuitive, which could lead the student to confusion. Then, the analysis is extended to general case, obtaining expressions applicable not only to the quarter wave line, but to lines of any length. In short, I will apply the transient analysis, actually presented in [5], in details to the quarter wavelength line.

It is also presented a simulation, using the LT spiceIV software [6], in which the transmission line was modeled as a cascade of series resistor and parallel inductors, in which all the results obtained were reproduced. This simulation may be of great interest for students, since the process of modeling the transmission line helps to obtain a deeper understanding of the basics concepts.

This work is written in a pedagogical style, and the principal aim is to serve as a complement for undergraduate students of a course on transmission line theory.

## II. EXAMPLE TO BE STUDIED AND CALCULATION OF COEFFICIENTS

It will be studied specifically the example given in [3], page 163, in which it is used a quarter wave transformer to match two transmission lines of characteristic impedance of 100 and 400 Ω. In this case, the characteristic impedance of the quarter wave transformer is: $Z_{0\lambda/4} = \sqrt{400 \times 100} = 200\Omega$. A schematic diagram for this case is shown at fig. 2.

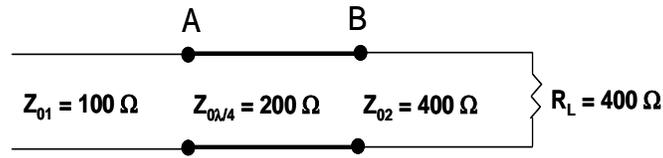

Figure 2. Schematic of the example studied in this work

The points marked as "A" and "B" at fig. 2 are the junctions, or interfaces (i.e., the points where the characteristics impedances change). The aim of the example is: giving a sin type continuous wave-front, of wavelength $\lambda$ and amplitude $V_0$, incident at point "A" from the left, determine the steady-state values of the amplitudes voltage waves to the left of point "A" and to the right of point "B". At these two points the reflections and transmission will be studied.

The transmission of the voltage wave at some interface is quantified trough the transmission coefficient for tension, defined as the ratio of the transmitted voltage amplitude to the incident voltage amplitude, and is given by [3]:

$$\tau = \frac{2Z_L}{Z_L + Z_0} = 1 + \rho. \tag{3}$$

Now, the calculations of reflection coefficients at interface "A" will be detailed:

$$\overleftarrow{\rho_A} = \frac{Z_{0\lambda/4} - Z_{01}}{Z_{0\lambda/4} + Z_{01}} = \frac{200 - 100}{200 + 100} = \frac{1}{3}. \tag{4}$$

The expression (1) was used, and the quarter wave line plays the role of "load". The arrow indicates the sense of propagation of the reflected wave. This fact must be considered because, at point "A", part of the wave will also be transmitted. This transmitted part will be partially reflected (and transmitted too) at interface "B". After that, the partial reflection coming from point "B" will be partially reflected at point "A". To take account quantitatively the reflection on "A" of a wave arriving from the right, the reflection coefficient is calculated as:

$$\overrightarrow{\rho_A} = \frac{Z_{01} - Z_{0\lambda/4}}{Z_{0\lambda/4} + Z_{01}} = \frac{100 - 200}{200 + 100} = -\frac{1}{3}. \tag{5}$$

Where, again, the arrow indicates the sense of propagation of the reflected wave. In this case, the 100 Ω line plays now the role of "load".

The transmissions coefficients are calculated using Eq. (3). The numeric results for reflections and transmission coefficients, for both possible reflections (and transmission) senses, at both interfaces "A" and "B" are summarized in table I.

TABLE I
REFLECTION AND TRANSMISSION COEFICIENT
AT BOTH INTERFACES "A" AND "B"

| Symbol | Quantity | Numerical Value |
|---|---|---|
| $\overleftarrow{\rho}_A$ | Reflection Coefficient | 1/3 |
| $\overrightarrow{\rho}_A$ | Reflection Coefficient | -1/3 |
| $\overrightarrow{\tau}_A$ | Transmission Coefficient | 4/3 |
| $\overleftarrow{\tau}_A$ | Transmission Coefficient | 2/3 |
| $\overleftarrow{\rho}_B$ | Reflection Coefficient | 1/3 |
| $\overrightarrow{\tau}_B$ | Transmission Coefficient | 4/3 |

Neither the coefficients $\overrightarrow{\rho}_B$ nor $\overleftarrow{\tau}_B$ were calculated, since it is not expected waves coming from the right at point "B", as the load $R_L$ is effectively matched with the 400 Ω line ($\overleftarrow{\rho} = 0$ at the resistor).

### III. TRANSIENT ANALYSIS AND CALCULATION OF STEADY-STATE VOLTAGE VALUES

Once the coefficients have been calculated, there will be analyzed the values of steady-state voltage values at each side of the quarter wave line. How is the steady-state value of the amplitude voltage wave of at the left of point "A" (i.e., at the 100 Ω line), calculated? This value will be called $V_{FA}$ and is expressed as a series. As there is a continuous wave-front with amplitude $V_0$ coming from the left, $V_0$ is the first term of the series. When this wave-front arrives at point "A", there will be a partial reflection, and the value of reflected amplitude voltage is $V_0 \overleftarrow{\rho}_A$, this being the second term of the series. At this point, there will be also a partial transmission of voltage wave, whose corresponding amplitude is $V_0 \overrightarrow{\tau}_A$. This wave, arriving at point "B" will be partially reflected and, returning to point "A", partially transmitted (and reflected too). The transmitted part, which is the third term of the series, is $V_0 \overrightarrow{\tau}_A \overleftarrow{\rho}_B \overleftarrow{\tau}_A e^{j2\beta\lambda/4}$, where the exponential term describes the phase shift of the voltage wave going from "A" to "B", and then returning all the way back. The factor $\beta = 2\pi/\lambda$ represents the propagation constant of the wave. The linear distance is, clearly, $2\lambda/4$. The reflected part is $V_0 \overrightarrow{\tau}_A \overleftarrow{\rho}_B \overrightarrow{\rho}_A e^{j2\beta\lambda/4}$, this wave will be also partially reflected and transmitted at "B" and "A", respectively, and contributes to the final value of $V_{FA}$ with the quantity $V_0 \overrightarrow{\tau}_A \overleftarrow{\rho}_B e^{j2\beta\lambda/4} \overrightarrow{\rho}_A \overleftarrow{\rho}_B \overleftarrow{\tau}_A e^{j2\beta\lambda/4}$. From this point on, the same reasoning must be applied, and the stationary value for the amplitude voltage at the left of "A" can be expressed as the infinite series:

$$V_{FA} = V_0 (1 + \overleftarrow{\rho}_A + \overrightarrow{\tau}_A \overleftarrow{\rho}_B \overleftarrow{\tau}_A e^{j2\beta\lambda/4} + \overrightarrow{\tau}_A \overleftarrow{\rho}_B e^{j2\beta\lambda/4} \overrightarrow{\rho}_A \overleftarrow{\rho}_B \overleftarrow{\tau}_A e^{j2\beta\lambda/4} + \overrightarrow{\tau}_A \overleftarrow{\rho}_B e^{j4\beta\lambda/4} \overrightarrow{\rho}_A \overleftarrow{\rho}_B \overrightarrow{\rho}_A \overleftarrow{\rho}_B \overleftarrow{\tau}_A e^{j2\beta\lambda/4} + ...). \quad (6)$$

A closer look to Eq. (6) reveals that, from the third term on, we have a common factor, and the series can be rewritten as:

$$V_{FA} = V_0 \{1 + \overleftarrow{\rho}_A + \overrightarrow{\tau}_A \overleftarrow{\rho}_B \overleftarrow{\tau}_A e^{j2\beta\lambda/4} [1 + \overrightarrow{\rho}_A \overleftarrow{\rho}_B e^{j2\beta\lambda/4} + (\overrightarrow{\rho}_A \overleftarrow{\rho}_B)^2 e^{j4\beta\lambda/4} + (\overrightarrow{\rho}_A \overleftarrow{\rho}_B)^3 e^{j6\beta\lambda/4} + ...]\}. \quad (7)$$

Where the term in brackets can be easily recognized as the geometric series, $1 + x + x^2 + x^3 + ..$, which converges to the value given by the expression $\frac{1}{1-x}$, provided that $|x|<1$. In our case, $x$ corresponds to $\overrightarrow{\rho}_A \overleftarrow{\rho}_B e^{j2\beta\lambda/4}$ and, since the reflections coefficients are lesser than one, the series converge and the expression (7) can be rewritten as:

$$V_{FA} = V_0 \left[1 + \overleftarrow{\rho}_A + \overrightarrow{\tau}_A \overleftarrow{\rho}_B \overleftarrow{\tau}_A e^{j2\beta\lambda/4} \left(\frac{1}{1 - \overrightarrow{\rho}_A \overleftarrow{\rho}_B e^{j2\beta\lambda/4}}\right)\right]. \quad (8)$$

Note that Eq. (8) represents a general expression, valid not only for the case of a quarter wavelength line, but for a line of arbitrary length, provided that instead of the $\lambda/4$ factor, the corresponding length (in units of $\lambda$) is used. For our particular case, the exponential term can be calculated as, $e^{2j\beta\lambda/4} = \cos\pi + j\sin\pi = -1$, where $\beta = 2\pi/\lambda$ was used, and, at last, the expression for $V_{FA}$ is:

$$V_{FA} = V_0 \left[1 + \overleftarrow{\rho}_A - \overrightarrow{\tau}_A \overleftarrow{\rho}_B \overleftarrow{\tau}_A \left(\frac{1}{1 + \overrightarrow{\rho}_A \overleftarrow{\rho}_B}\right)\right]. \quad (9)$$

Using the values for the calculated coefficients listed in Table I, the value of $V_{FA}$ can be calculated, the result is $V_{FA} = V_0$, which shows that the reflections at the interface "A", are gradually cancelled and, in stationary state, quarter wave line effectively matches the impedances, as expected.

The first term in Eq. (8) represents the amplitude of the incident wave, the others terms describe the amplitude of the reflected wave. With the definition of the reflection coefficient, Eq. (1), and rewriting Eq. (8) as:

$$V_{FA} = V_0[1 + \overleftarrow{\rho_{AE}}]. \tag{10}$$

The second term of Eq. (10), $\overleftarrow{\rho_{AE}}$, can be understood as an effective reflection coefficient at point "A". The expression for this coefficient is obtained by comparing Eq. (10) and Eq. (8):

$$\overleftarrow{\rho_{AE}} = \overleftarrow{\rho_A} + \overrightarrow{\tau_A}\overleftarrow{\rho_B}\overleftarrow{\tau_A} e^{j2\beta\lambda/4}\left(\frac{1}{1 - \overrightarrow{\rho_A}\overleftarrow{\rho_B} e^{j2\beta\lambda/4}}\right). \tag{11}$$

Of course, the value of $\overleftarrow{\rho_{AE}}$, for the case of quarter wave line, is zero.

Note in Eq. (11), that the reflection coefficient of voltage, defined in Eq. (1), is just the first term of an infinite series.

Now, it will be calculated the steady-state value of the amplitude voltage wave to the right of interface "B", that is, in the 400 Ω line (and also at the load $R_L$). This voltage will be called $V_{FB}$. To get this result, the same reasoning as before must be employed. The answer will be again expressed as an infinite series.

As already seen, the voltage wave, with amplitude $V_0$ arriving at point "A" from the left, will be partially transmitted (with amplitude $V_0\overrightarrow{\tau_A}$), and then partially transmitted at interface "B", with amplitude $V_0\overrightarrow{\tau_A}\overrightarrow{\tau_B}$, which is the first term of the series. To obtain the second term, we must calculate, from amplitude $V_0\overrightarrow{\tau_A}$, the reflected amplitude voltage at interface "B", then, the reflected voltage at "A", and finally, the transmitted voltage at "B", taking account the phase shift. As a result for the second term of the series, it is obtained $V_0\overrightarrow{\tau_A}\overleftarrow{\rho_B}\overrightarrow{\rho_A}\overrightarrow{\tau_B} e^{j2\beta\lambda/4}$. The other terms can be obtained in a similar way, giving:

$$V_{FB} = V_0(\overrightarrow{\tau_A}\overrightarrow{\tau_B} + \overrightarrow{\tau_A}\overleftarrow{\rho_B}\overrightarrow{\rho_A}\overrightarrow{\tau_B} e^{j2\beta\lambda/4} + \overrightarrow{\tau_A}\overleftarrow{\rho_B}\overrightarrow{\rho_A} e^{j2\beta\lambda/4}\overleftarrow{\rho_B}\overrightarrow{\rho_A}\overrightarrow{\tau_B} e^{j2\beta\lambda/4} + \overrightarrow{\tau_A}\overleftarrow{\rho_B}\overrightarrow{\rho_A} e^{j4\beta\lambda/4}\overleftarrow{\rho_B}\overrightarrow{\rho_A}\overleftarrow{\rho_B}\overrightarrow{\rho_A}\overrightarrow{\tau_B} e^{j2\beta\lambda/4} + ...) . \tag{12}$$

It is observed in Eq. (12), a common factor and, after rewriting, the expression above can be expressed as:

$$V_{FB} = V_0\{\overrightarrow{\tau_A}\overrightarrow{\tau_B} + \overrightarrow{\tau_A}\overrightarrow{\tau_B}\overleftarrow{\rho_B}\overrightarrow{\rho_A} e^{j2\beta\lambda/4}[1 + \overrightarrow{\rho_A}\overleftarrow{\rho_B} e^{j2\beta\lambda/4} + (\overrightarrow{\rho_A}\overleftarrow{\rho_B} e^{j2\beta\lambda/4})^2 + (\overrightarrow{\rho_A}\overleftarrow{\rho_B} e^{j2\beta\lambda/4})^3 + ....]\}, \tag{13}$$

where, once again, the geometric series is recognized and finally, Eq. (13) is written as:

$$V_{FB} = V_0 \overrightarrow{\tau_A}\overrightarrow{\tau_B}[1 + \overleftarrow{\rho_B}\overrightarrow{\rho_A} e^{j2\beta\lambda/4}(\frac{1}{1 - \overrightarrow{\rho_A}\overleftarrow{\rho_B} e^{j2\beta\lambda/4}})] . \tag{14}$$

Also, Eq. (14) can be used as a general expression for calculating $V_{FB}$, substituting the term λ/4 at the exponential factor by the length (in λ units), of the intermediate line.

For the quarter wavelength line, the exponential term is -1, as seen before, and the result for the steady-state voltage value is: $V_{FB} = 2V_0$. This is coherent with the result presented in [3].

From Eq. (14), it is possible to write the expression for an effective transmission coefficient at point "B":

$$\overrightarrow{\tau_{BE}} = \overrightarrow{\tau_A}\overrightarrow{\tau_B}[1 + \overleftarrow{\rho_B}\overrightarrow{\rho_A} e^{j2\beta\lambda/4}(\frac{1}{1 - \overleftarrow{\rho_B}\overrightarrow{\rho_A} e^{j2\beta\lambda/4}})] \tag{15}$$

which, for this particular case, takes the value 2.

### IV. SIMULATIONS

In order to reproduce the results obtained, a simulation using LT Spice IV [6] was performed. As a font of the sin-type voltage waves, it was used an electrical generator, with a frequency, $f$, of 13,4 Mhz. To simulate the transmission line, a cascade of cells consisting of inductors (connected in series) and capacitors (connected in parallel) were assembled, as shown in the schematic in Fig. 3.

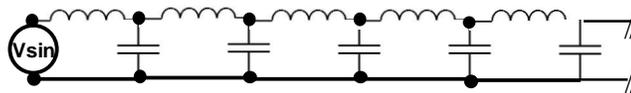

Figure 3. Transmission line as simulated in LT SPICE IV.

The characteristics impedance of the line, $Z_0$ is given by:

$$Z_0 = \sqrt{\frac{L}{C}}. \qquad (16)$$

Where $L$ and $C$ are the inductance and capacitance per unit of length, respectively. In the particular case simulated here, it was emulated the transmission line with a $Z_0 = 100 \, \Omega$, connecting forty six cells, each consisting in a series inductance of 0.53 µH and a parallel capacitance of 52.1 pF. In the context of the simulation, this values of inductance and capacitance are actually the values of inductance and capacitance per "unit of length", which can be defined as the longitudinal length of one cell. Assuming this unit of length to be meter, the value of $L$ and $C$ should be chosen also in order to keep the velocity of propagation of the voltage waves to a value physically realistic. This velocity is given by:

$$v = \sqrt{\frac{1}{LC}}. \qquad (17)$$

With the values of L and C used in this example, the velocity calculated using (17) is $1.91 \times 10^8$ m/s, which is a realistic value for a transmission line. Of course, it can be assumed another units of length for the individual cells, and the values of $L$ and $C$ should have the appropriate values, in order that (17) continues to give a physically realistic value of $v$. With the values of $v$ and $f$, the wavelength, $\lambda$, of the tension wave propagating on this line can be calculated trough the well known relation:

$$\lambda = v/f, \qquad (18)$$

and has the value $\lambda \cong 14.3$ m. Now, the quarter wave line must be designed. As already seen, the characteristic impedance for this line must be $Z_{0\lambda/4} = 200 \, \Omega$. The values of inductance and capacitance per unit of length for this line were chosen to be $L_{\lambda/4} = 0.93$ µH/m and $C_{\lambda/4} = 23.32$ pF/m, respectively. To calculate the appropriate length of this line, the value of $\lambda_{\lambda/4}$ must be known. It can be calculated by first estimating the velocity, using the values of $L_{\lambda/4}$ and $C_{\lambda/4}$ in (17), and then using (18). The result is $\lambda_{\lambda/4}$ ~16 units of length, or 16 m. Then, to emulate the quarter wavelength line in our simulation, there must be placed four cells of inductances and capacitances with the values $L_{\lambda/4}$ and $C_{\lambda/4}$.

In order to not complicate the simulation, the line of $Z_0 = 400 \, \Omega$, can be replaced by the load, $R_L = 400 \, \Omega$, with no difference in the final results.

It was simulated a temporal window of 700 ns. The results of the simulation are shown in fig. 4.

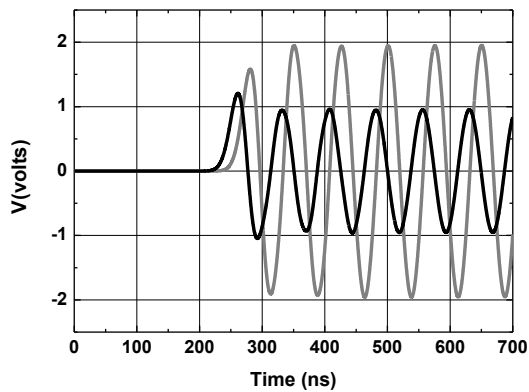

Figure 4. Voltage waves obtained in simulation. Black trace: Voltage at point A(see Fig. 2); Gray trace: Voltage at point B.

The voltage waves shown in Fig. 4 were obtained at point "A", in black trace, and at point "B", in gray trace (see Fig.2). Note the delays of ~220 ns and 250 ns for both the traces, indicating the time the waves took to travel from the generator to points "A" and "B", respectively. It also can be seen, in Fig 4, that the amplitude voltage values obtained in the simulation are slightly lesser than expected, i.e., 1V and 2 V for traces black and gray, respectively. This is due to the fact that the inductor used in the software has a default series resistance of 1 mΩ, which produce this potential drop.

## V. CONCLUSIONS

In this work, there have been determined the steady-state values of the amplitude voltage waves at the extremes of a quarter wave line, for an incident voltage sin type wave-front with amplitude $V_0$ and frequency $f$. An apparent contradiction, due to the fact that coefficient of refection for voltage is not zero, when calculated using Eq. (1) at the interface "A", was resolved using a transient analysis.

The expressions obtained can be used as generics formulas, for the case of transmission lines of arbitrary length making a junction between two other lines, or between a line and a load, just substituting, in Eq. (8) and Eq. (14), the λ/4 factor by the length of the intermediary line.

There where also obtained expressions for the effective transmission and reflection coefficients for voltage at the two extremes of the quarter wavelength line. These expressions can also be used for lines of arbitrary lengths.

The numeric results obtained were reproduced using the LT Spice IV software, were the transmission lines were simulated as a cascaded of cells consisting inductors and capacitors.

The results here presented are not generally detailed in elementary textbooks on transmission line for courses of electrical engineering or applied physics. It is expected that this work can be used as a complement by the students when studying these topics.

## REFERENCES

[1] F. A. Callegari, "Análise dos transientes de ondas de tensão em ambos os extremos de uma linha quarto de onda e analogia com o fenômeno de reflexão e transmissão de ondas eletromagnéticas", Revista Brasileira de Ensino de Física, vol. 33, N⁰ 2, 2011.
[2] Ulaby, F.T., *Electromagnetics for Engineers*, Prentice Hall, Upper Saddle River, New Jersey, 2005.
[3] J. D. Krauss, and D. A. Fleisch. *Electromagnetics with Applications*. WCB/McGraw-Hill, 1999.
[4] S. M. Wentworth, Fundamentals of Electromagnetics with Engineering Applications, Wiley, 2006.
[5] Robert E. Collin, *Foundations for microwave engineering*, McGraw-Hill Book Company, New York, 1966.
[6] http://www.linear.com/designtools/software/